\documentclass[aps,pra,reprint,showpacs,superscriptaddress]{revtex4-1}

\usepackage{amsmath}
\usepackage{amsfonts}
\usepackage{stmaryrd}
\usepackage{amssymb}
\usepackage{graphics,epsfig}
\usepackage{subfigure}
\usepackage{bm}
\usepackage{mathpazo}
\usepackage{CJK}

\usepackage[dvipdfm,
            pdfstartview=FitH,
            CJKbookmarks=true,
            bookmarksnumbered=true,
            bookmarksopen=true,
            linktocpage=true,
            colorlinks=true, 
            pdfborder=001,   
            citecolor=blue,  
            urlcolor=blue,   
            linkcolor=blue,  
            anchorcolor=blue,
            ]{hyperref}      

\begin{document}

\begin{CJK*}{GBK}{kai}

\title{Probing the Conformations of Single Molecule via Photon Counting Statistics}

\author{Yonggang Peng }
\affiliation{School of Physics, Shandong University, Jinan 250100, China}

 \author{Zhen-Dong Sun }
  \affiliation{School of Physics, Shandong University, Jinan 250100, China}

 \author{Chuanlu Yang }
 \affiliation{School of Physics and Optoelectronic Engineering, Ludong University, Yantai 264025, China}

  \author{Yujun Zheng }
  \email{Electronic mail: yzheng@sdu.edu.cn}
  \affiliation{School of Physics, Shandong University, Jinan 250100, China}

\begin{abstract}
We suggest an approach to detect the conformation of
    single molecule by using the photon counting statistics.
    The generalized Smoluchoswki equation is employed to describe
    the dynamical process of conformational change of single molecule.
    The resonant trajectories of the emission photon numbers $\langle N \rangle$
    and the Mandel's $Q$ parameter, in the space of conformational
    coordinates $\bm{\mathcal{X}}$ and frequency $\omega_L$ of external
    field ($\bm{\mathcal{X}}-\omega_L$ space), can be used to rebuild the
    conformation of the single molecule. As an example, we consider
    Thioflavin T molecule. It demonstrates that the results of conformations extracted
    by employing the photon counting statistics is excellent agreement with
    the results of {\it ab initio} computation.
\end{abstract}







\maketitle

  \end{CJK*}

 \section{Introduction}

    The single molecule technique,  excluding the ensemble average
    compared with traditional techniques, can be used to detect and
    measure the dynamical process in the level of single molecule~\cite{barkai2004}.
    This technique could help us to discover some new quantum phenomena
    occurred in single molecule level, such as, spectral diffusion,
    fluorescence intermittency~\cite{zheng2003prl,barkai2004} {\it etc.}.
    With the advancement of the experimental techniques, the  single
    molecule technique has become one of the useful techniques for
    studying physical, chemical and material science.
    However, there are some difficulties in the single molecule experiments.
    One of the major difficulties is
    the conformational dynamics strongly effects the single
    molecule signals under different conditions.

    The conformational information of single molecule
    plays an important role in physical, chemical and
    biological processes. For example, the conformational changes of some
    organic molecules can affect the efficient of the solar
    cells~\cite{xie,gao1,gao2}. Also, neurodegenerative Alzheimer's
    and Parkinson's diseases, cataracts associate with unnormal conformation of
    the macro-biomlecules~\cite{selkoe2003,zerovnik2002,harper1997,carrell1998,koo1999,guo2013}.
    Several methods are developed to study the conformational changes
    of single molecules experimentally and theoretically, such as,
    single pair fluorescence resonance energy transfer (sp-FRET)~\cite{weiss2000},
    single molecule fluorescence polarization anisotropy (sm-FPA),
    and atomic force microscopy~\cite{agnihotri2004}, molecular dynamics
    simulation~\cite{voelz2010,Liu2012}, Monte Carlo simulation~\cite{senderowitz1995},
    and kinetic method~\cite{presse2013,neuweiler2010,agmonjpcb2000,zhengjcp2004,
    JPhysChemB.110.19040,JPhysChemB.112.12867} {\it etc.}.
    In these methods, it is assumed that the single molecule transits between some discrete
    states for investigating the single molecule conformational dynamics.
    In the simulation and the kinetic methods,
    the transition rate constants between different states
    are obtained via the best fitting experimental data. However,
    in some cases, the conformational changes of single molecule show
    complex behaviors since the conformational changes of single molecule
    depend on its conformational coordinates $\mathcal{X}$.
    The methods mentioned above are hardly to completely describe
    the conformational changes with the case that the single molecule is
    changing with the conformational coordinates $\mathcal{X}$.

To overcome this inadequacy, one develops different methods to
simulate the conformations
numerically, and compares the numerical simulation conformation
with in-direct experimental results. The question, however, is that
can we direct probe the conformations of single molecule experimentally,
and how can we do this? As we know, the emission photons from the
single molecule not only include the information of the molecule
interaction with its surrounding environments but also include the
information of the single molecule itself. Such as, the spectral diffusion
process of the single molecule reflects the dynamics of the single
molecule interaction with its surrounding
 molecules~\cite{zheng2003prl,zheng20031,JChemPhys.139.164120,barkai2004},
and the blinking phenomenon in the single quantum dots reflects the
details of quantum process which occurred in the systems.
Experimentally, the factorial moments of emitted photons $\langle N_r \rangle$,
arrival times and frequencies of emitted photons {\it etc.} are serving as
the experimental data, which 
contain information about the nature of conformations and time scale of
the underlying conformational changes.
In this paper, we suggest an approach to extract the
conformational information of single molecule in the space of the conformational
coordinate and the frequency of external field ($\bm{\mathcal{X}}-\omega_L$ space)
from the resonant trajectory of emitted photon. As a concrete example,
we take the single molecule of Thioflavin T to demonstrate the
application of our theoretical approach extracting its conformation numerically.

 \section{Theoretical framework}

We assume a single molecule has several energy levels $\{|n(\bm{\mathcal{X}})\rangle\}$,
and its corresponding eigen-energy is $\epsilon_n(\bm{\mathcal{X}})$,
where $\bm{\mathcal{X}}$ is the conformational coordinate.
In the energy picture of the single molecule, $\epsilon_n(\bm{\mathcal{X}})$ can also
be thought as the ``energy surface''. The conformational dynamics of  the
single molecule can be thought as the diffusion in different energy surfaces.
It hence could be assumed the single molecular system satisfies
the generalized Smoluchoswki equation
\begin{eqnarray}
\label{eq:gen-smo}
\frac{\partial}{\partial t}\rho(\bm{\mathcal{X}},t) &=&
        -\frac{i}{\hslash}[\mathcal{H},\rho(\bm{\mathcal{X}},t)]
        +\mathcal{Z}\rho(\bm{\mathcal{X}},t) \nonumber\\ &&
        +\mathcal{R}\rho(\bm{\mathcal{X}},t)+\mathcal{L}\rho(\bm{\mathcal{X}},t),
\end{eqnarray}
where $\mathcal{H}$ is the Hamiltonian of the single molecular system,
including the Hamiltonian of ``bare'' single molecule
and the interaction between the single molecule and external field,
$\mathcal{Z} \rho$ describes the conformational ``diffusion'' process of
the single molecule, $\mathcal{R}\rho$ describes the
transitions caused by the system interacting with it surrounding
environments, and $\mathcal{L}\rho$ describes the
spontaneous emission process. $\mathcal{R}$ and $\mathcal{L}$ are
the environment assistant transition and
spontaneous emission operators, respectively~\cite{JPhysChemB.110.19066}.

The operator $\mathcal{Z}$ is the Smoluchowski operator, it can,
via its acting on density matrix, be defined as
\begin{eqnarray}
\left( \mathcal{Z}\rho \right)_{nm} &=& \frac{1}{2} \frac{\partial}{\partial \bm{\mathcal{X}}}
      D e^{-\epsilon_n(\bm{\mathcal{X}})}
      \frac{\partial}{\partial \bm{\mathcal{X}}} e^{+\epsilon_n( \bm{\mathcal{X}})}\rho_{nm} + \nonumber\\
      & & \frac{1}{2} \frac{\partial}{\partial \bm{\mathcal{X}}} D e^{-\epsilon_m(\bm{\mathcal{X}})}
      \frac{\partial}{\partial \bm{\mathcal{X}}} e^{+\epsilon_m(\bm{\mathcal{X}})}\rho_{nm},
\end{eqnarray}
where $D$ is the diffusion coefficient.

The Langevin equation can be employed to determine the surface
$\epsilon_1(\bm{\mathcal{X}})$ of single molecule in ground state.
The surface of single molecule in ground state  can be expressed as
\begin{equation}
   \epsilon_1(\bm{\mathcal{X}}) = \epsilon_1(\bm{\mathcal{X}}_{e})+
   \frac{1}{D} \int_{\bm{\mathcal{X}}}^{\bm{\mathcal{X}}_{e}} \dot{\bm{r}}_1 d\bm{r}_1 ,
\end{equation}
where $\bm{\mathcal{X}}_{e}$ is the equilibrium position.
For convenient, we can assume $ \epsilon_1(\bm{\mathcal{X}}_{e})=0$.

The single molecule can, using the precise tuning capability, be pumped
by scanning microscopy. It is experimentally imaging the resonant
excitation and photon emission. That is, we can
obtain resonant peaks from the emitted photon counting statistics of
first factorial moment $\langle N \rangle$ or the second factorial moment of
the Mandel's $Q$ parameter for long time limit.
Each of the resonant peaks of photon counting statistics corresponds to
an resonant absorption between the different conformational excited
states and ground state of single molecule in $\bm{\mathcal{X}}-\omega_L$ space.

If we denote $\hslash\omega_{mn}^{(r)}(\bm{\mathcal{X}})$ as
the resonant absorption peak between the states $|m\rangle$ and $|n\rangle$
in $\bm{\mathcal{X}}-\omega_L$ space. Or,
we define $\omega_{mn}^{(r)}(\bm{\mathcal{X}})$ as
{\it a resonant trajectory} of emitted photon for long time
limit in $\bm{\mathcal{X}}-\omega_L$ space,
we have
\begin{equation}
  \hslash\omega_{mn}^{(r)} (\bm{\mathcal{X}}) = \epsilon_m(\bm{\mathcal{X}})-
           \epsilon_n(\bm{\mathcal{X}}).
\end{equation}

Then, one can obtain the conformation of $\epsilon_m(\bm{\mathcal{X}})$
from the resonant trajectory $\hslash\omega_{mn}^{(r)}(\bm{\mathcal{X}})$ and
the known conformation of $\epsilon_n(\bm{\mathcal{X}})$.
The conformation of $\epsilon_m(\bm{\mathcal{X}})$ can be
expressed as following
\begin{equation}
 \label{eq:traj}
 \epsilon_m(\bm{\mathcal{X}}) = \hslash\omega_{mn}^{(r)}(\bm{\mathcal{X}})+
        \epsilon_n (\bm{\mathcal{X}}),
\end{equation}
by employing {\it a resonant trajectory}.

 \section{An Example}

We numerically demonstrate that {\it the resonant trajectories} of
photon counting moments are employed to extract the conformational changes of
the single Thioflavin T (ThT) molecule. The fluorescence intensity change of
the ThT molecule is considered associating with its conformational changes,
which can be used to detect the dynamics of the amyloid fibrils~\cite{levine1993}.
In this paper, we consider the case that the conformation of ThT molecule
is changed with a torsion angle $\varphi$, which is the angle between
the benzthiazole and the dimethylaminobenzene rings (for ThT molecule, the
conformational changes can also be thought as configurational changes).
Namely, in this case we have $\bm{\mathcal{X}}=\varphi$.
The ThT molecule can be described by four levels~\cite{stsiapura2007,stsiapura2008,stsiapura2010}:
two conformational ground states $|1\rangle$ and $|2\rangle$,
and two conformational excited states $|3\rangle$ and $|4\rangle$.
The transitions between the ground state $|1\rangle~ (|2\rangle)$
to the excited states $|3\rangle$ and $|4\rangle$
are dipole transition allowed. The transition between confromational
ground (excited) states $|2\rangle$ and $|1\rangle$
($|4\rangle$ and $|3\rangle$) is dipole transition forbidden.

The Hamiltonian of the single molecule can be expressed as
\begin{equation}
\label{eq:ham}
\mathcal{H}=\mathcal{H}_0+\mathcal{H}',
\end{equation}
where $\mathcal{H}_0$, the Hamiltonian of the ``bare'' single molecule,
can be expressed as
\begin{equation}
\label{eq:ham0}
        \mathcal{H}_0 = \sum_{n=1}^{4} \epsilon_n(\varphi) |n
        \rangle\langle n|,
\end{equation}
where $\epsilon_n(\varphi)=\hslash\omega_n(\varphi)$ is the eigen-energy.
The interaction between the single molecule and the external field reads
\begin{equation}
\label{eq:ham-int}
\mathcal{H}' = -\sum_{mn} \bm{\mu}_{mn}\cdot \mathbf{\mathcal{E}}_0(t) \cos(\omega_L t) (a_{mn}^{\dag}+a_{mn}),
\end{equation}
where  $a_{mn}^{\dag}=|m\rangle\langle n|$,
and $a_{mn}=|n\rangle\langle m|$, with $m=3,4$ and $n=1,2$,
$\bm{\mu}_{mn}$ is the transition dipole between the states $|m\rangle$
and $|n\rangle$, $\mathcal{E}_0(t)$ and $\omega_L$ are
the amplitude and the angular frequency of the external field, respectively.

Based on Eqs.~(\ref{eq:ham0}) and~(\ref{eq:ham-int}),
the Hamiltonian~(\ref{eq:ham}) of the single molecule in
the representation of the ``bare'' single molecule $\mathcal{H}_0$ and
under the rotating wave approximation (RWA), can be written as
\begin{equation}
    \mathcal{H}(\varphi)= \hslash \left(
		\begin{array}{cccc}
			0 & 0 & \Omega_{31}(t)/2 & \Omega_{41}(t)/2\\
			0 & \omega_{21} & \Omega_{32}(t)/2 & \Omega_{42}(t)/2\\
			\Omega_{31}(t)/2 & \Omega_{32}(t)/2 & -(\Delta+\omega_{43}) & 0\\
			\Omega_{41}(t)/2 & \Omega_{42}(t)/2 & 0 & -\Delta
		\end{array}
	\right),
\end{equation}
where $\Delta=\omega_L - \omega_{41}$ is the detuning frequency between external
field $\omega_L$ and the transition frequency $\omega_{41}=\omega_4(\varphi)-\omega_1(\varphi)$,
$\Omega_{nm}(t)=-\bm{\mu}_{nm}\cdot \mathcal{E}_0(t)/\hslash$ is the Rabi frequencies.

To obtain the resonant absorption trajectory of photon counting
moments of $\langle N \rangle (\varphi, \omega_L)$ and the Mandel's
$Q(\varphi,\omega_L)$ parameter in the $\varphi-\omega_L$ space,
the generating function approach is employed
to simulate the trajectories of $\langle N \rangle (\varphi, \omega_L)$ and
$Q(\varphi,\omega_L)$  for long time limit theoretically.
As noted in previous works \cite{zheng2003prl,zhengjcp2004,barkaiprl,
 JPhysChemB.110.19066,zheng20031,JChemPhys.139.164120,hepra,heprl,mukamelpra},
 the message taken with the emitted photon for long time
limit from single molecules is
the information when the single molecule is being excited.
Hence, we could note down the scenario of the conformation of single molecule on
the time of being excited using the resonant trajectory in
$\bm{\mathcal{X}}-\omega_L$ space for long time limit via the generating
function approach of photon counting statistics.
We can scan the single molecule to obtain the resonant trajectory in
$\bm{\mathcal{X}}-\omega_L$ space by employing the (ultra)fast laser (we mean
here the laser is faster than the conformational dynamics of single molecule) to
pump the single molecule.

The generating function is defined as follows~\cite{zheng2003prl,zheng20031,
JChemPhys.139.164120,heprl,mukamelpra},
\begin{equation}
\label{eq:def-gen}
    \mathcal{G}(\varphi,s,t)=\sum_n \rho^{(n)}(\varphi,t) s^n,
\end{equation}
where $\sum_{n=0}^\infty \rho^{(n)}(\varphi,t)=\rho(\varphi,t)$.
$\rho^{(n)}(\varphi,t)$ is the partition of single molecule system emitted $n$ photons
at the time interval $[0,t]$ in the conformation $\varphi$,
$s$ is the auxiliary parameter for counting photons.
The generating function, based on the generalized Smoluchoswki
equation~(\ref{eq:gen-smo}) and after some algebra using Eq.~(\ref{eq:def-gen}),
satisfies the following equation (The exciting process is great faster than
the conformational dynamics by using ultra-fast laser field. Here, we omit
the conformational operator in our equation.)
\begin{eqnarray}
\label{eq:density}
\frac{\partial}{\partial t}\mathcal{G}(\varphi,s,t) &=&
 -\frac{i}{\hslash} \left[\mathcal{H}(\varphi,t),\mathcal{G}(\varphi,s,t) \right]
      +\sum_{m,n} (\mathcal{Z}\mathcal{G}(\varphi,s,t))_{mn} \nonumber\\
   & & + \sum_{m,n} \frac{\Gamma_{mn} }{2} \left( 2 s a_{mn} \mathcal{G}(\varphi,s,t) a_{mn}^{\dag} - \right. \nonumber\\
    & & \left.  a_{mn}^{\dag} a_{mn} \mathcal{G}(\varphi,s,t) -
       \mathcal{G}(\varphi,s,t) a_{mn}^{\dag} a_{mn} \right)  \nonumber\\
    & & + \sum_{m,n}\frac{\mathcal{R}_{mn}}{2} \left(2 a_{mn}\mathcal{G}(\varphi,s,t) a_{mn}^{\dag} - \right. \nonumber\\
    && \left.   a_{mn}^{\dag} a_{mn} \mathcal{G}(\varphi,s,t) -\mathcal{G}(\varphi,s,t) a_{mn}^{\dag} a_{mn} \right),
\end{eqnarray}
where $\Gamma_{mn}$ is the spontaneous emission rate
between the states $|m\rangle$ and $|n\rangle$,
$\mathcal{R}_{mn}$ is the environment assistant transition rate
between the states $|m\rangle$ and $|n\rangle$.

The factorial moments of $\langle N_r \rangle$ can be simply obtained by
taking derivatives with respect to $s$ evaluated at $s=1$, namely
\begin{eqnarray}
    \langle N_r \rangle (\varphi,t) &\equiv& \langle N(N-1)(N-2) \cdots (N-r+1) \rangle \\\nonumber
           &=& \left. \frac{\partial^r}{\partial s^r}\mathcal{Y}(\varphi,s,t) \right|_{s=1},
\end{eqnarray}
and the Mandel's $Q(\varphi,t)$ parameter calculated as
\begin{equation}
    Q(\varphi,t)=\frac{\langle N_2 \rangle(\varphi,t)-\langle N_1 \rangle^2(\varphi,t)}
             {\langle N_1 \rangle(\varphi,t)} ,
\label{eq:q}
\end{equation}
where the working generating function $\mathcal{Y}(\varphi,s,t)$ is defined as follows
\begin{equation}
    \mathcal{Y}(\varphi,s,t)=\sum_{n=1}^4 \mathcal{G}_{nn}(\varphi,s,t).
\end{equation}

Scanning the single molecule system, we can obtain the resonant trajectories of
the mean number of photons $\langle N \rangle(\varphi,\omega_L) =\langle N_1 \rangle(\varphi,\omega_L)$
and $Q(\varphi,\omega_L)$ for long time $t$ in $\varphi - \omega_L$ space.

In our numerical simulations, the spontaneous emission rate $\Gamma_{mn}$ is
existence for $m=3,4$ and $n=1,2$; and only the environment assistant transition
rates $\mathcal{R}_{21}$ and $\mathcal{R}_{43}$ are existence.
    For ThT molecule, the conformational dynamics (the corresponding time scale
    $\tau_c \sim 10^{-11} - 10^{-9}$ s) is great faster than the spontaneous emission
    transition (the corresponding time scale $\tau_f \sim 10^{-8}$ s). One can employ
    an ultra-short laser pulse to excite the single molecule, and the width of the laser
    pulse $T$ would less shorter than conformational dynamical time scale $\tau_c$. As
    we know, a finite length laser wave could cause the laser frequency has a boarden
    $\delta\omega\sim 1/T$. If $\delta\omega$ is larger or the same as the energy gap
    between two excited states or ground states, the states can not be distinguished.
    The width of laser pulse $T$ can be chosen as $10^{-14}\le T \le 10^{-11}$, and
    the pulse area $\Theta_{mn} = \int_{-\infty}^{\infty} \Omega_{mn}(t) d t \gg \pi$,  which
    could promise the exciting process is great faster than conformational dynamics.
In our calculation, we choose
    the pulse area $\Theta_{mn}=100\pi$.

   \begin{figure}[ht]
        \includegraphics[width=3in]{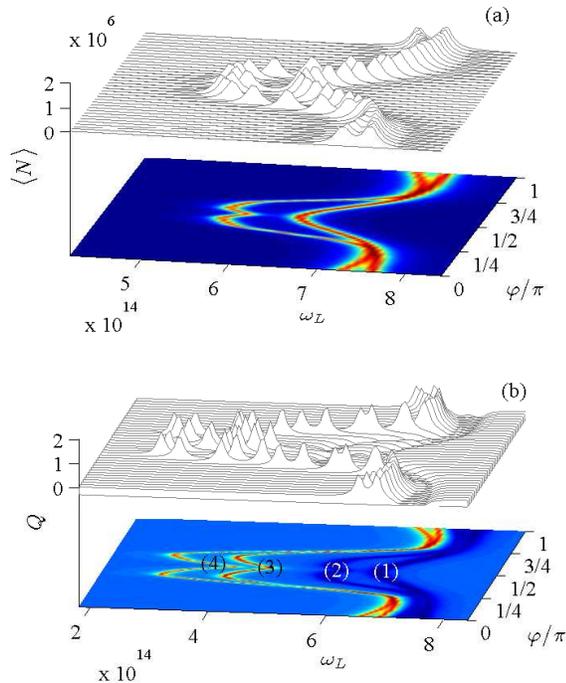}
      \caption{The resonant trajectories of emission photon numbers $\langle N \rangle$
      and the Mandel's $Q$ parameter in the conformational coordinate and frequency space
    (torsion angle $\varphi$ and the laser frequency $\omega_L$).
    The Rabi frequency $\Omega_{14}=\Omega_{24}=\Omega_{13}=\Omega_{23}=10^4\gamma$,
    the spontaneous emission rates $\Gamma_{41}=\Gamma_{42}=\Gamma_{31}=\Gamma_{32}=\gamma$,
    and environment assistant transition rates
    $\mathcal{R}_{21}=\mathcal{R}_{43}=\gamma$.
}
    \label{fig:emission}
    \end{figure}

Figure~\ref{fig:emission} numerically simulates the possible experiment
results of the resonant trajectories of emission photons $\langle N \rangle$
and the Mandel's $Q$ parameter in the $\varphi - \omega_L$ (conformational-frequency) space.
The spontaneous emission rates $\Gamma_{41}=\Gamma_{42}=\Gamma_{31}=\Gamma_{32}=\gamma=3\times 10^8$ Hz,
the Rabi frequency $\Omega_{14}=\Omega_{13}=\Omega_{24}=\Omega_{23}=10^4\gamma$
and the environment assistant transition rates $\mathcal{R}_{21}=\mathcal{R}_{43}=\gamma$~\cite{note3},
which corresponds to relaxation between the excited states $\epsilon_4(\varphi)$
and $\epsilon_3(\varphi)$ and the ground states $\epsilon_2(\varphi)$ and $\epsilon_1(\varphi)$.
  The resonant trajectories of the emission photons
  $\langle N \rangle(\varphi,\omega_L)$ and the Mandel's
  $Q (\varphi,\omega_L)$ in Fig.~\ref{fig:emission} correspond to the resonant absorption
  transition between the single molecule energy levels.
  In the Mandel's $Q(\varphi,\omega_L)$ panel
  of Fig.~\ref{fig:emission},
  trajectory $1$ corresponds to the resonance transition
  $|4\rangle \to  |1\rangle$;
  trajectory $2$ corresponds to the transition $|3\rangle \to |1\rangle$;
  trajectory $3$ corresponds to the transition $|4\rangle \to |2\rangle$;
  trajectory $4$ corresponds to the transition $|4\rangle \to |2\rangle$.
  That means (we denote the curves frequency by
  $\omega_{tn}(\varphi), (n=1,2,3,4)$) $\omega_{t1}(\varphi)
  =\omega_4(\varphi)-\omega_1(\varphi)$,
  $\omega_{t2}(\varphi)=\omega_3(\varphi)-\omega_1(\varphi)$,
  $\omega_{t3}(\varphi)=\omega_4(\varphi)-\omega_2(\varphi)$,
  and $\omega_{t4}(\varphi)=\omega_3(\varphi)-\omega_2(\varphi)$.
  Then, we can extract the conformational information of the single molecule:
   $\omega_4(\varphi)=\omega_{t1}(\varphi)+\omega_1(\varphi)$,
  $\omega_3(\varphi)=\omega_{t2}(\varphi)+\omega_1(\varphi)$, and
  $\omega_2(\varphi)=\omega_{t1}(\varphi)-\omega_{t3}(\varphi)+\omega_1(\varphi)$.

  \begin{figure}
  \includegraphics[width=3in]{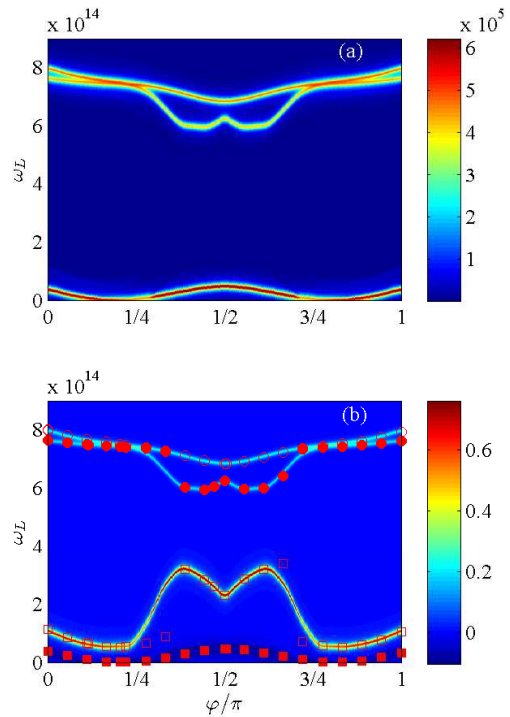}
    \caption{(Color Online) The conformational structures of the ThT single molecule.
    The conformation of the single Th T molecule is
    recreased by the resonant trajectories in the conformational coordinate and
    frequency space shown in Fig.~\ref{fig:emission}.
    The marks are the results of {\it ab initio}.\cite{stsiapura2007}}
    \label{fig:recreat}
  \end{figure}

  Figure~\ref{fig:recreat} demonstrates the numerical results of
  the conformations of the ThT single molecule which is extracted
  using the resonant trajectories (shown in Fig.~\ref{fig:emission}) of
  the emission photons $\langle N \rangle(\varphi,\omega_L)$
  (a) and Mandel's $Q(\varphi,\omega_L)$ parameter (b).

  It should be noted that, as shown in the top panel of Fig.~\ref{fig:recreat},
  we can not obtain the surface $\epsilon_2(\varphi)$ of single molecule
  from the resonant trajectory of
  the mean number of photons $\langle N \rangle(\varphi,\omega_L)$.
   The reason is that the surface of state $|2\rangle$ is
  more unstable than state $|1\rangle$ for the ThT molecule,
  the population of state $|2\rangle$ is
  less than that of state $|1\rangle$, and the probability of
  the system transition from state $|2\rangle$ into states $|3\rangle$
  and $|4\rangle$ is very small. This results in the resonant trajectories
  $\langle N \rangle(\varphi,\omega_L)$ of the transition
  from state $|2\rangle$ into states $|3\rangle$ and $|4\rangle$
  are very weak (see top panel of Fig.~\ref{fig:emission}).
   However, the resonant trajectory of the Mandel's $Q$ parameter,
   related to the second factorial moment,
   as another probing signal is strong enough. As shown in the bottom panel of
   Fig.~\ref{fig:emission}, the resonant trajectory of the Mandel's $Q$ parameter
  represents all the conformational structures.
  As the comparison, we also show the results of  conformation of ThT
  by employing {\it ab initio} simulations of Ref.~\cite{stsiapura2007},
  the results of {\it ab initio} simulations are marked using circles, solid-circles, squares and
  solid-squares, respectively. The results are excellent agreement with each other.

 \section{Concluding remark}

In this paper, we demonstrate an approach obtaining single molecule conformational structures via
photon counting statistics. The Langevin and generalized Smoluchoswki equations are employed
to describe the conformational dynamics. The resonant trajectories of the emission
photon numbers $\langle N \rangle$ and Mandel's $Q$ parameter in conformational-frequency space
$(\mathcal{X}-\omega_L)$ include all the conformational information of single molecule. The
single molecule conformational structures can be obtained from the resonant trajectory in
$(\mathcal{X}-\omega_L)$ space. The ThT molecule as an example is demonstrated, and the results
are excellent agreement with that of \emph{ab initio}.

\begin{acknowledgments}
The authors thank Frank Brown for useful discussions.
This work was supported by the National Natural Science
Foundation (Grand Nos. 11374191,11074147, 11174186) and the National
Basic Research Program of China (973 Program, Grant
No. 2015CB921004).
\end{acknowledgments}


\begin{thebibliography}{36}
\bibitem [{not()}]{note3}%
  \BibitemOpen
  \href@noop {} {}\bibinfo {note} {The transition rates $\mathcal{R}_{21}$ and
  $\mathcal{R}_{43}$ can be considered the single molecule colliding with
  solution molecules. The collision life time $\tau_0$ can be expressed as
  $\frac{1}{\tau_0} = 4 d^2 n \sqrt{\frac{\pi k_B T}{M}}$~\cite{loudonbook},
  where $d$ is the distant between the molecules, $n$ is the molecule number
  density, $k_B$ is Boltzman's constant, $T$ is the temperature, and $M$ is the
  molecular mass. For the single molecule experiments at room temperature, the
  density $n \in (1, 10^6)$ n mol/L~\cite{selvinbook}. The corresponding life
  time $\tau_0 \in (10^{-3}, 10^{-9})$ s. In our calculation, we choose
  $1/\tau_0=3\times 10^{8}$ Hz.}\BibitemShut {Stop}%
\end{thebibliography}
\end{document}